\documentclass[copyright,creativecommons,noncommercial]{eptcs}
\providecommand{\event}{TFPIE 2025} % Name of the event you are submitting to

\usepackage{iftex}

\ifpdf
  \usepackage{underscore}         % Only needed if you use pdflatex.
  \usepackage[T1]{fontenc}        % Recommended with pdflatex
\else
  \usepackage{breakurl}           % Not needed if you use pdflatex only.
\fi

\usepackage{amsmath,amssymb,stmaryrd}
\usepackage{mathtools}
\usepackage{listings}
\usepackage[scaled=0.92]{helvet}

\usepackage{amsthm}
\theoremstyle{remark}
\newtheorem{example}{Example}
\theoremstyle{plain}
\newtheorem{theorem}{Theorem}

\usepackage{xcolor}
\colorlet{new}{black}%{magenta}
\newcommand\NEW[1]{{\color{new}#1}}
\newenvironment{ENEW}{\color{new}}{}

\newcommand\Dist{\mathcal{D}}
\newcommand\NaN{\mathrm{NaN}}
\newcommand\aware{\stackrel\star}
\newcommand\absy[1]{\ensuremath{\mathit{#1}}}
\newcommand\infrule[2]{\dfrac{#1}{#2}}
\newcommand\sem[2][]{[\kern-1.5pt[#2]\kern-1.5pt]_{\mathrm{#1}}}
\newcommand\shortmapsto{\mathbin{\mapstochar\shortrightarrow}}

\usepackage{xcolor}

\title{{\scshape alea iacta est}: A Declarative Domain-Specific Language for Manually Performable Random Experiments}
\author{Baltasar Trancón y Widemann
\institute{semantics gGmbH, Berlin, DE}
\institute{Brandenburg University of Applied Sciences, DE}
\email{trancon@th-brandenburg.de}
\and
Markus Lepper
\institute{semantics gGmbH, Berlin, DE}
}
\def\titlerunning{\scshape alea iacta est}
\def\authorrunning{B.\ Trancón y Widemann \& M.\ Lepper}

\hypersetup{pdftitle={alea iacta est: A Declarative Domain-Specific Language for Manually Performable Random Experiments},pdfauthor={B.\ Trancón y Widemann \& M.\ Lepper}}

\begin{document}
\maketitle

\begin{abstract}
  Random experiments that are simple and clear enough to be performed
  by human agents feature prominently in the teaching of elementary
  stochastics as well as in games.  We present \textsf{Alea}, a
  domain-specific language for the specification of random
  experiments.  \textsf{Alea} code can either be analyzed statically
  to obtain and inspect probability distributions of outcomes, or be
  executed with a source pseudo-randomness for simulation or as a game
  assistant.  The language is intended for ease of use by non-expert
  programmers, \NEW{in particular students of elementary stochastics,
    and players and designers of games of chance,} by focusing on
  concepts common to functional programming and basic mathematics.
  Both the design of the language and the implementation of runtime
  environments are work in progress.
\end{abstract}

\section{Introduction}
\label{intro}

Human culture has devised many ways for performing controlled simple
random experiments: Bones, dice and coins can be thrown, distinct
items can be drawn from bags, urns or shuffled card decks, etcetera.
The outcomes have been used to drive games of chance, simulations and
sometimes even serious decision making.  The mathematical field of
stochastics has, to a substantial degree, arisen from the analysis of
such experiments \cite{DeMoivre1718}.  Even though it has later been
expanded to encompass vastly more abstract and powerful notions of
uncertainty, the simple experiments still have a prominent place in
its teaching.

The most basic experiments consist of isolated random operations:
\begin{example}
  \label{example:denarius}
  \begin{quote}
    \itshape Toss a denarius coin.  Determine whether it shows a ship or
    a head.\;\footnote{An ancient Roman game \cite{coin}.}
  \end{quote}
\end{example}

More interesting experiments are created by aggregating random data, and
by performing transformations and statistical calculations on them:
\begin{example}
  \label{example:vampire}
  \leavevmode
  \begin{quote}
    \itshape Roll seven ten-sided dice.  For every die that shows a
    ten, roll another and add it to the pool.  Count the number of
    dice that show values greater than five.  Compare to the number of
    dice that show a one.  If the difference is positive, you win by
    that amount.  If the difference is zero or negative, you lose.  If
    there are no values greater than five but there is a one, you lose
    badly.\;\footnote{A rule from the tabletop RPG \emph{Vampire: The
        Masquerade} \cite{vampire}.}
  \end{quote}
\end{example}

\textsf{Alea} is being designed as a notation that gives precise
syntax and semantics to such random experiments.  As the examples
indicate, a functional language with rich data types but simple
control flow, orthogonally extended with randomness, is called for.
Ideally, the language should be intuitive to use \NEW{for expressing
  simple mathematical formulas}, without advanced programming skills,
but still offer the benefits of a `real' programming language in terms
of automatic checking and interpretation.  It should be expressive
enough that random experiments can be specified clearly and concisely,
but not too powerful for exhaustive static analysis.

A teaching aid could take the form of a `computer stochastics system'
akin to a `computer algebra system', with interactive analysis and
evaluation of expressions, and means for exploring the outcomes
visually and computationally.  Game assistant functionality could
range from simple `digital dice', with the additional features of
automatic evaluation of outcomes and visibility control in
multi-player games, to a `chance oracle' for calibrating the odds of
individual game situations, or even overall game rules, a priori.

\section{Design of Semantic Domain}

The design of \textsf{Alea} is founded on semantic requirements that
impose limits on the admissible expressive power of language features.

\subsection{Probability}

The semantics of \textsf{Alea} expressions shall be
\emph{stochastically effective}: Any well-formed and well-typed
expression must be associated with a unique probability distribution
that can be computed systematically, without recourse to algebraic
meta-knowledge or heuristics.

\begin{ENEW}
  
  We only consider probability distributions that are \emph{simple} in
  the following sense: Let $X$ be an arbitrary set.  Write
  $\mathcal{D}(X)$ for the rational-valued, finitely supported
  probability mass functions on $X$, i.e., $P \in \mathcal{D}(X)$ if
  and only if:
  \begin{align*}
    P &: X \to \mathbb{Q} & P(x) &\geq 0 &
    \sum_{x \in X} P(x) &= 1 & \operatorname{supp}(P)\enspace\text{finite}
  \end{align*}
  Each such function $P$ completely defines a distribution on $X$,
  which is consequently called \emph{rational}, \emph{discrete}, and
  \emph{finite}.
  
  The inner product of a simple distribution with any other
  number-valued function is a finite sum, even if the function domain
  $X$ is infinite:
  \begin{align*}
    \sum_{\mathclap{x \in X}} P(x) \cdot f(x) =
    \sum_{\mathclap{x \in \operatorname{supp}(P)}} P(x) \cdot f(x)
  \end{align*}

  Hence the \emph{mean} of a simple distribution over the rationals
  $P : \mathbb{Q} \to \mathbb{Q}$ is straightforwardly computable:
  \begin{align*}
    m(P) = \sum_{\mathclap{x \in \operatorname{supp}(P)}} P(x) \cdot x
  \end{align*}

\end{ENEW}

The construct $\Dist$ can be given the rich category-theoretic
structure of a \emph{commutative}, or \emph{symmetric monoidal monad};
\NEW{see \cite[e.g.]{DBLP:journals/jlp/Jacobs18}}.  Namely, it gives
rise to a functor that generalizes computing the distribution of
random variables $f$ from the distribution of underlying events,
\begin{align*}
  \Dist(f : X \to Y) &: \Dist(X) \to \Dist(Y) &
  \Dist(f)(P)(y) &= \sum_{x \in X} P(x) \cdot [f(x) = y]
\end{align*}
with the Kronecker delta distribution as the monad unit,
\begin{align*}
  \delta_X &: X \to \Dist(X) & \delta_X(x)(x') &= [x = x']
\end{align*}
the folding of two layers of a probability tree into one as the monad
multiplication,
\begin{align*}
  \mu_X &: \Dist\bigl(\Dist(X)\bigr) \to \Dist(X) &
  \mu_X(P)(x) = \sum_{Q \in \Dist(X)} P(Q) \cdot Q(x)
\end{align*}
and the stochastically independent combination of marginal distributions,
\begin{align*}
  \psi_{X,Y} &: \Dist(X) \times \Dist(Y) \to \Dist(X \times Y) &
  \psi_{X,Y}(P, Q)(x, y) = P(x) \cdot Q(y)
\end{align*}
which are all natural and satisfy several coherence laws.  These
properties allow us to `lift' the intuitively evident meaning of a
deterministic term language to a stochastic reading in a systematic
way, by means of Moggi-style \cite{MOGGI199155} denotational
semantics.

\begin{quote}
  For example, consider an expression of the form $f(a, b)$, where $a$
  and $b$ are subexpressions that evaluate deterministically to the
  values $v_a \in A$ and $v_b \in B$, respectively, and
  $f : A \times B \to C$ is a known binary function.  Then the overall
  value is clearly $f(v_a, v_b) \in C$.  Now let $b$ (but not $a$)
  instead evaluate stochastically to a distribution
  $P_b \in \Dist(B)$, and $f$ be also of the stochastic type
  $f : A \times B \to \Dist(C)$.  Then the overall distribution
  is given canonically by
  $\mu_C\bigl(\Dist(f)\bigl(\psi_{A,B}(\delta_A(v_a),
  P_b)\bigr)\bigr) \in \Dist(C)$.
\end{quote}

Furthermore, all distributions have finite representations and all
monadic operations are inherently computable.  Hence, if used together
with computable random variables only, the distribution of any
finitary expression is computable and the equality of distributions is
decidable, in a constructive way.

On the downside, the class $\Dist$ is not closed under unbounded
iteration or recursion:
\begin{example}
  \label{example:geom}
  \begin{quote}
    \textit{``Repeat rolling a die until it shows a six, counting the
      number of rolls''}
  \end{quote}
\end{example}
This experiment does \emph{not} have a simple distribution, in the
above sense.  Hence, as the core of a programming language, the
calculus of simple distributions is not Turing-complete by itself, and
can be embedded in, but not fully integrated with a Turing-complete
language without losing its beneficial properties.  We have decided to
accept that limitation on the expressivity of \textsf{Alea}, at least
in the current version of the language.

\subsection{Data}

The intuition underlying the data model of \textsf{Alea} is discussed
here briefly and informally.  For a more structural treatment, see
section~\ref{types} below.

\subsubsection{Primitive Data}

Most primitive data involved in random experiments are either numeric
(for arithmetic computation) or of a finite enumerated type (for
classification).  \textsf{Alea} supports the rational numbers and
their subalgebras, the integers and naturals.  Enumerated types are
supported by tagging values with symbolic identifiers, in the sense of
algebraic data type constructors.

A special role is played by the Boolean truth values, which could be
understood as either numeric or enumerated.  Since it appears that the
majority of uses of Boolean values in specifications of random
experiments is for \emph{counting}, \textsf{Alea} treats them
literally as the values $\{ 0, 1 \}$, a subtype of the natural
numbers.  Note that for languages that treat the Booleans as
non-numerical, the \emph{Iverson bracket} $[ \_ ]$ that maps them to
their number counterparts is a ubiquitous operation in counting
computations.

The set of numeric values is extended with the special value $\NaN$
(\emph{not-a-number}) with semantics analogous to the standard for
floating-point numbers \cite{IEEE754}: Undefined arithmetic
operations, most notably division by zero, yield $\NaN$.  If any
operand is $\NaN$, arithmetic operations also yield $\NaN$, whereas
comparison operations yield false.\footnote{With the notable exception
  of $\neq$.}  While this convention is well-understood and works fine
for purely numerical data, interesting problems arise when aggregate
data are considered; see below.

\subsubsection{Collections}

Multiple data elements of the same type may be aggregated into
collections.  In addition to the most popular collection shapes in
functional programming, \emph{lists} and \emph{sets}, \textsf{Alea}
provides built-in support for \emph{multisets} aka \emph{bags}.  Bags
should be preferred over lists in any situation where the ordering of
elements is irrelevant, and comes with two significant benefits:
\begin{enumerate}
\item The number of possible outcomes, and hence the size of
  distributions, may be reduced dramatically, by as much as $n!$, if
  subsequent interpretation is only up to permutation.
\item The bulk processing with \emph{map}, \emph{filter} and
  \emph{reduce} operations can be accelerated whenever an individual
  element occurs many times in a bag; to a single step for \emph{map}
  and \emph{reduce}, and to a logarithmic number of steps for
  \emph{reduce} with an associative binary operation, by means of the
  exponentiation-by-squaring algorithm.
\end{enumerate}

In all other aspects, the treatment of all three collection shapes in
the language is as similar as possible.

The semantics of collections are complicated by the fact that the
object-level comparison operator $\aware=$ for numerical
values\footnote{Written as \texttt{=} in the front-end notation.} is
different from the meta-level set-theoretic identity relation $=$;
namely $\NaN \aware= \NaN$ is false.  For sets in \textsf{Alea}
semantics, we specify that
\begin{equation*}
  S \aware\subseteq T \iff
  \forall x \in S\mathpunct. \exists y \in T\mathpunct. x \aware= y
\end{equation*}
and derive:
\begin{align*}
  S \aware\supseteq T &\iff T \aware\subseteq S
  &
  S \aware\subset T &\iff S \aware\subseteq T \land \lnot(S \aware\supseteq T)
  &
  S \aware= T &\iff S \aware\subseteq T \land S \aware\supseteq T
  \\
  &&
  S \aware\supset T &\iff S \aware\supseteq T \land \lnot(S \aware\subseteq T)
  &
  S \aware\neq T &\iff \lnot(S \aware= T)
\end{align*}

\begin{quote}
  For example, we find that
  $\{ 2, 3 \} \aware\subset \{ 1, 2, 3, \NaN \}$ is true, but
  $\{ \NaN \} \aware= \{ \NaN \}$ is false.
\end{quote}

The construction can be transferred to bags by accounting for
multiplicity.  For lists, there are several contenders for a `sublist'
relationship.  We currently assume that `contiguous sublist anywhere'
is the most useful one and define list comparison accordingly, for
instance $[2, 3] \aware\subset [1, 2, 3, 4]$ but
$[1, 4] \aware{\not\subset} [1, 2, 3, 4]$; albeit that design choice
is somewhat contentious.

\begin{ENEW}

  \subsection{Related Work}

  There is a great body of work on the general theory and the specific
  design of \emph{probabilistic} programming languages; see
  \cite{Barthe_Katoen_Silva_2020} for a detailed book-length overview.
  However, we have found that the requirements discussed above do not
  have a perfect match in any of the available theoretical frameworks.
  For context, we review some common topics briefly, even though they
  may not apply directly to \textsf{Alea}.

  \subsubsection{Probabilistic Control Flow}

  Imperative programming is first of all about control flow.  Hence,
  probabilistic imperative languages need to deal with
  probabily-weighted branching and looping
  \cite{Barthe_Katoen_Silva_2020_1,Kozen1985}; even termination may be
  probabilistic.  By contrast, \textsf{Alea} is concerned with data
  flow, and its programs terminate by construction.

  \subsubsection{Statistical Inference}

  Probabilistic domain-specific description languages are used to
  specify statistical models in the context of Bayesian inference
  \cite{Barthe_Katoen_Silva_2020_2}.  Such languages tend to
  incorporate distributions outside the class $\Dist$, in particular
  continous ones such as exponential or beta distributions, as well as
  general recursion and imprecise floating-point arithmetics.
  Consequently, their semantics are inherently based on sampling and,
  possibly quite sophisticated, Monte Carlo estimation methods
  \cite[e.g.]{Goodman2008,Wood2014}.  By contrast, \textsf{Alea} guarantees
  exactly computable distributions for all expressions.

  \subsubsection{Computability in the Limit}

  In theory, the rather restricted class of computable distributions
  $\Dist$ can be extended by relaxing the notion of computability to
  accomodate (computable approximations of) real numbers
  \cite{Weihrauch2000}, and by extension, continuous distributions
  \cite{Barthe_Katoen_Silva_2020_3}.  For fundamental reasons, the
  equality predicate $=$ on the reals is not computable in such a
  framework.  By contrast, \textsf{Alea} provides set-theoretic
  semantics, including equality as a primitive, to the programmer.

  \subsubsection{Recursive Distributions}

  A more modest extension is the combination of discrete exact
  distributions with recursion \cite{Barthe_Katoen_Silva_2020_4}.
  This admits simple infinite distributions, such as the geometric
  distribution arising in Example~\ref{example:geom}, in a natural
  way.  They are given semantics analogous to ordinary recursive
  functions, which is fitting in a calculus of higher-order functions.
  By contrast, \textsf{Alea} is first-order, and makes distributions
  analogous to other kinds of data; in fact, it comes with an
  \emph{expectation} operator that works exactly like a \emph{reduce}
  operation on a collection.
  
\end{ENEW}

\section{Notation and Concepts}

\subsection{Syntax}
\label{notation}

The \emph{concrete} front-end syntax of \textsf{Alea} is not yet
finalized.  Hence we will not give a formal grammar here, but describe
the overall style by examples.  By contrast, see Section~\ref{eval}
for an \emph{abstract} syntax with all due semantic formalities.

\begin{ENEW}
  The principal design guideline is that both syntax and semantics
  should follow the conventions of elementary mathematics as far as
  possible, more importantly than those of programming in general, or
  any particular style of programming languages in particular.
\end{ENEW}

\subsubsection{Numbers and Arithmetics}

\textsf{Alea} features a single unified type of numbers, namely
unlimited-precision rationals, with all the usual arithmetic and
comparison operators.
\begin{lstlisting}[gobble=4,xleftmargin=4em]
  % 3 * (x + 1) << -2/3
\end{lstlisting}
The special value $\absy{NaN}$ can be written as \lstinline{0/0}.
Integer and natural numbers are recognized as subtypes, and come with
integer division operators \lstinline{//}~(\emph{div}) and
\lstinline{\\}~(\emph{mod}).

Booleans are just the numbers zero (\emph{false}) and one
(\emph{true}).  Hence the logical operators $\wedge$ and $\vee$ are
just synonyms for \emph{min} and \emph{max}, respectively.  Unlike in
many languages where numbers and Booleans are distinct, $+$ and $*$
with one or more Boolean operands are useful for counting and
numerical filtering, respectively.

\subsubsection{Collection Notation}

\textsf{Alea} supports three shapes of collections, namely lists, bags
and sets.  In the syntax, they are mostly distinguished by the brace
shapes \lstinline{[ ]}, \lstinline{< >} (or \lstinline{<! >!})
and \lstinline|{ }|, respectively.  Collections can be created by
enumerating the elements or specifying ranges.
\begin{lstlisting}[gobble=4,xleftmargin=4em]
  % [1, 2, 3, 4, 5]        <2, 2, 4, 3, 2>        {1 .. 5}
\end{lstlisting}

Collections are processed using bulk operations.  The equivalent of
the usual higher-order functions \emph{map} and \emph{filter} is
expressed in first-order notation with Haskell-style comprehensions
with a \emph{map} clause of the left hand side, and \emph{generator}
and \emph{filter} clauses on the right hand side of the vertical bar.
\begin{lstlisting}[gobble=4,xleftmargin=4em,escapechar=!]
  % { k*k | k <- {1 .. n}; even(k) }
  %  !\rmfamily\itshape\footnotesize(map)!    !\rmfamily\itshape\footnotesize(generator)!     !\rmfamily\itshape\footnotesize(filter)!
\end{lstlisting}

See \cite{Gibbons2016} for a truly `comprehensive' introduction to the
semantical background.

Comprehensions for bags and lists have analogous notations enclosed by
their respective style of brackets.  Generator clauses draw elements
from source collections of either the same shape as the target
collection, or one that is naturally convertible:
\begin{itemize}
\item Lists are naturally convertible to bags, by forgetting order.
\item Bags are naturally convertible to sets, by forgetting
  multiplicity.
\item Lists are naturally convertible to sets, by forgetting both.
\end{itemize}

If there is no explicit \emph{map} clause on the left hand side of the
vertical bar, the leftmost generator clause can take its place,
leading to a style that closely resembles traditional set-theoretical
notation.
\begin{lstlisting}[gobble=4,xleftmargin=4em]
  % { x <- S | x >= 0 }    =    { x | x <- S; x >= 0 }    /=    { x >= 0 | x <- S }
\end{lstlisting}

The combination of the two preceding features allows for a concise
notation for natural conversion; for instance, a list \lstinline{L}
can be converted to a set by writing \lstinline!{ x <- L }!, or even
\lstinline!{ _ <- L }!.

Generator clauses are right-associative; \lstinline{x <- y <- z} is an
abbreviation for \lstinline{y <- z; x <- y} (note the transposition).
Thus the monad multiplication, or \emph{flatten} operation, on a
collection \lstinline{C} is conveniently expressed as
\lstinline{[ x <- y <- C ]}, or even \lstinline{[ _ <- _ <- C ]}.

Unusually for a programming language with comprehensions,
\textsf{Alea} also supports `drawing without replacement'.
\begin{lstlisting}[gobble=4,xleftmargin=4em]
  % { b - a | {a, b} <- {1 .. n}; a << b }
\end{lstlisting}

The equivalent of the usual higher-order function \emph{reduce} is
expressed by applying a binary function to a single collection
argument.  Operators can be used as functions in this sense, but must
be quoted in parentheses.  Furthermore, the function must be known to
\dots
\begin{itemize}
\item be associative,
\item be commutative if the collection is a bag or set,
\item be idempotent if the collection is a set,
\item have a neutral element if the collection may be empty.
\end{itemize}
How to declare or check these semigroup-theoretic properties is
out of scope here.
\begin{lstlisting}[gobble=4,xleftmargin=4em]
  % max{x, y, z}        (+)< k >= 0 | k <- B >
\end{lstlisting}
The latter example demonstrates how summation over a bag of Booleans,
resulting in a natural number, is conveniently used for counting.
(In this example, the nonnegative elements of \lstinline{B} are counted.)

\subsubsection{Named Functions and Distributions}

\textsf{Alea} comes with a library of predefined functions, which are
named but not conceptually different from operators.  The argument
must be given in parentheses, unless it is already some sort of
bracketed expression.
\begin{lstlisting}[gobble=4,xleftmargin=4em]
  % max(a, b) + min{x, y, z}
\end{lstlisting}

Additionally, the library contains predefined distributions, which
are syntactically marked with \lstinline{~}, and may also take dynamic
parameters.  The invocation of a distribution can be thought of as an
anonymous random variable.  Note that \textsf{Alea} is not
referentially transparent in this respect; every evaluation of a
distribution expression is stochastically independent from the other.
The expectation operator is written as the function \lstinline{E}.
\begin{lstlisting}[gobble=4,xleftmargin=4em]
  % E(~uniform{1 ..  6} + ~bernoulli(2/3))
\end{lstlisting}

\subsubsection{Records}

Records, aggregate data with several fields of heterogeneous type, are
written in parentheses.  The fields can be identified explicitly by
name or implicitly by position.  Tuples are just records with only
positional fields.  Mixing both styles is permitted, but not
recommended.  Selection of fields from a record is written with the
usual dot notation.  As in Standard ML, positional field numbers start
from one.  They must be marked with \lstinline{#} to avoid ambiguity
with decimal fractions.
\begin{lstlisting}[gobble=4,xleftmargin=4em]
  % (foo: 42, bar: {})        (1, 2, 3)        r.foo        s.#3
\end{lstlisting}

\subsubsection{Tagged Values}

Tagged values are elements of sum types.  The tag names, both as
constructors and in patterns, are written like function applications
but marked with \lstinline{@}.  If the argument is omitted, it
defaults to the empty tuple \lstinline{()}.  Thus, tags also serve as
enumerated constants.  Tag can be used freely, without having to
declare which sum type they inject into.  Tagged data are processed
using elementary pattern matching in the style of Haskell (core).
\begin{lstlisting}[gobble=4,xleftmargin=4em]
  % @good(42)        @bad        x ? { @good(n) ->  n; @bad ->  -1 }
\end{lstlisting}

Explicit matching cases must be non-overlapping; an additional default
case is written with the pattern \lstinline{_}.  The same notation is also used for a C-style switch statement on numerical values.
\begin{lstlisting}[gobble=4,xleftmargin=4em]
  % n ? { 0 -> @none; 1 ->  @one; 2, 3 -> @few; _ -> @many }
\end{lstlisting}

The special case of Boolean values can be abbreviated, also in a C-like style.
\begin{lstlisting}[gobble=4,xleftmargin=4em]
  % b ? { 1 -> x; 0 -> y }        b ? x : y
\end{lstlisting}

\subsubsection{Non-Features}

\textsf{Alea} does not have many beloved features of typical
functional programming languages: No recursion, no partially defined
operations such as list indexing, no higher-order functions, and no
lambdas.  It is a deliberate design choice that the expression
notation uses only graphical symbols, no keywords.

\subsection{Type System}
\label{types}

The type universe of \textsf{Alea} is specified by the following
abstract syntax:
\begin{align*}
  \begin{aligned}
    \absy{Type} ::= {}& \absy{any} \mid \absy{none}
    \\
    \mid {}& \absy{num}(\absy{Num})
    \mid \absy{coll}(\absy{Shape}, \absy{Mode}, \absy{Type})
    \\
    \mid {}& \absy{prod}(\absy{FieldId} \nrightarrow \absy{Type})
    \mid \absy{sum}(\absy{CaseId} \nrightarrow \absy{Type})
  \end{aligned}
  &&
  \begin{aligned}
    \absy{Num} &::= \absy{bool} \mid \absy{nat} \mid \absy{int} \mid \absy{rat}
    \\
    \absy{Shape} &::= \absy{list} \mid \absy{bag} \mid \absy{set}
    \\
    \absy{Mode} &::= \absy{pos} \mid \absy{opt}
  \end{aligned}
\end{align*}

Note that this abstract syntax is designed for extensibility,
orthogonality and conceptual simplicity of computation, not for ease
of notation in the front-end language.  However, this is not a
pressing problem, because types feature only implicitly in end-user
code.

\begin{itemize}
\item The types $\absy{any}$ and $\absy{none}$ denote the upper and
  lower bound of the type lattice, respectively.
\item The constructor $\absy{num}$ denotes numeric types.
\item The constructor $\absy{coll}$ denotes collection types, which
  come in the shapes $\absy{list}$, $\absy{set}$ and $\absy{bag}$, are
  of the mode $\absy{pos}$ (positive, may not be empty) or
  $\absy{opt}$ (optional, may be empty) and have a specific element
  type.
\item The constructor $\absy{prod}$ denotes product types, which have
  a finite set of orthogonal fields specified as a mapping of field
  identifiers to field types.  Field identifiers can be symbolic or
  positional, such that tuples are merely a special case of products.
\item The constructor $\absy{sum}$ denotes sum types, which have a
  finite set of disjoint cases specified as a mapping of case
  identifiers to case types.  Enumerated types are special cases of
  sum types, where every case type is the unit type
  $\absy{prod}(\varnothing)$.
\end{itemize}

\subsubsection{Semantics of Types}

Since here functions, let alone recursive ones, are not values that would
need to be typed, a simple set-theoretic semantics can be given that
assigns to every type its \emph{extension}.  Let there be an untyped value
universe specified by the following abstract syntax:
\begin{align*}
  \absy{Val} ::= {}& \absy{const}(\mathbb{Q}) \mid \absy{NaN}
  \mid \absy{thelist}\bigl(\mathbf{L}(\absy{Val})\bigr)
  \mid \absy{thebag}\bigl(\mathbf{M}(\absy{Val})\bigr)
  \mid \absy{theset}\bigl(\mathbf{P}(\absy{Val})\bigr)
  \\
  \mid {}& \absy{record}(\absy{FieldId} \nrightarrow \absy{Val})
  \mid \absy{tag}(\absy{CaseId} \times \absy{Val})
\end{align*}

For the sake of uniformity, we write the operators $\mathbf{L}$,
$\mathbf{M}$ and $\mathbf{P}$ for the finite lists, bags, and subsets
of a set, respectively, and $\varnothing$ for the empty list, bag and
set.  Note that we assume that number spaces are properly embedded:
$\mathbb{B} \subset \mathbb{N} \subset \mathbb{Z} \subset \mathbb{Q}$.

The extensional semantics of types is a map
$\sem[V]{\_} : \absy{Type} \to \mathbf{P}(\absy{Val})$, the details are given in
Figure~\ref{fig:sem-v}.

\begin{figure}
  \centering
  \fbox{\begin{minipage}{.9\textwidth}
      \begin{align*}
        \sem[V]{\absy{none}} &= \varnothing
        &
        \sem[V]{\absy{any}} &= \absy{Val}  
      \end{align*}
      \begin{align*}
        \sem[V]{\absy{num}(\absy{bool})} &= \absy{const}(\mathbb{B})
        &
        \sem[V]{\absy{num}(\absy{nat})} &= \absy{const}(\mathbb{N}) \cup \{ \absy{NaN} \}
        \\
        \sem[V]{\absy{num}(\absy{int})} &= \absy{const}(\mathbb{Z}) \cup \{ \absy{NaN} \}
        &
        \sem[V]{\absy{num}(\absy{rat})} &= \absy{const}(\mathbb{Q}) \cup \{ \absy{NaN} \}
      \end{align*}
      \begin{align*}
        \sem[V]{\absy{coll}(\absy{list}, \absy{opt}, t)} &= \absy{thelist}\bigl(\mathbf{L}(\sem[V]{t})\bigr)
        &
        \sem[V]{\absy{coll}(\absy{list}, \absy{pos}, t)} &= \absy{thelist}\bigl(\mathbf{L}(\sem[V]{t}) \setminus \{ \varnothing \}\bigr)
        \\
        \sem[V]{\absy{coll}(\absy{bag}, \absy{opt}, t)} &= \absy{thebag}\bigl(\mathbf{M}(\sem[V]{t})\bigr)
        &
        \sem[V]{\absy{coll}(\absy{bag}, \absy{pos}, t)} &= \absy{thebag}\bigl(\mathbf{M}(\sem[V]{t}) \setminus \{ \varnothing \}\bigr)
        \\
        \sem[V]{\absy{coll}(\absy{set}, \absy{opt}, t)} &= \absy{theset}\bigl(\mathbf{P}(\sem[V]{t})\bigr)
        &
        \sem[V]{\absy{coll}(\absy{set}, \absy{pos}, t)} &= \absy{theset}\bigl(\mathbf{P}(\sem[V]{t}) \setminus \{ \varnothing \}\bigr)
      \end{align*}
      \begin{align*}
        \sem[V]{\absy{prod}(T)} &= \absy{record}\bigl(\bigl\{ f : \absy{FieldId} \nrightarrow \absy{Val} \bigm| \forall i \in \operatorname{dom}(T)\mathpunct. f(i) \in \sem[V]{T(i)} \bigr\}\bigr)
        \\
        \sem[V]{\absy{sum}(T)} &= \bigcup_{i \in \operatorname{dom}(T)} \absy{tag}\bigl(\{ i \} \times \sem[V]{T(i)}\bigr)
      \end{align*}
      \vspace{0pt}
    \end{minipage}}
  \caption{Extensional Type Semantics}
  \label{fig:sem-v}
\end{figure}

\subsubsection{Empty and Inhabited Types}

The following inference rules deduce that a type is empty on a
syntactical basis:
\begin{align*}
  \infrule{}{\operatorname{empty} \absy{none}} && \infrule{\forall i \in \operatorname{dom}(T)\mathpunct. \operatorname{empty} T(i)}{\operatorname{empty} \absy{sum}(T)}
\end{align*}
\begin{align*}
  \infrule{\operatorname{empty} t}{\operatorname{empty} \absy{coll}(s, \absy{pos}, t)}
  &&
  \infrule{\exists i \in \operatorname{dom}(T)\mathpunct. \operatorname{empty} T(i)}{\operatorname{empty} \absy{prod}(T)}
\end{align*}

\begin{ENEW}
  
  \begin{theorem} The emptiness rules are sound and
    complete:\;\footnote{\NEW{Formal proofs of this and subsequent
        theorems are not interesting enough by themselves to be given
        here; they belong into a technical report.}}
    \begin{align*}
      \operatorname{empty} t \iff \sem[V]{t} = \varnothing
    \end{align*}
  \end{theorem}

  \begin{proof}
    By induction.
  \end{proof}

\end{ENEW}

\subsubsection{Subtyping}

\textsf{Alea} types form a lattice.  The inference rules of the
subtyping calculus are given in Figure~\ref{fig:subtype}.  The
presentation has been geared towards highlighting how independent
sublattices are pieced together.  As a result, some statements are a
little indirect; most notably $s_1 \sqsubseteq s_2$ is merely saying
$s_1 = s_2$, since there are no collection subshapes, at least not in
the current version of the type system.

\begin{figure}
  \centering
  \fbox{\begin{minipage}{.9\textwidth}
      \begin{align*}
        \infrule{}{\absy{none} \sqsubseteq t \sqsubseteq \absy{any}}
        &&
        \infrule{n_1 \sqsubseteq n_2}{\absy{num}(n_1) \sqsubseteq \absy{num}(n_2)}
        &&
        \infrule{}{\absy{bool} \sqsubseteq \absy{nat} \sqsubseteq \absy{int} \sqsubseteq \absy{rat}}
      \end{align*}
      \begin{align*}
        \infrule{s_1 \sqsubseteq s_2 \quad m_1 \sqsubseteq m_2 \quad t_1 \sqsubseteq t_2}{\absy{coll}(s_1, m_1, t_1) \sqsubseteq \absy{coll}(s_2, m_2, t_2)} && \infrule{}{\absy{pos} \sqsubseteq \absy{opt}}
      \end{align*}
      \begin{align*}
        \infrule{\forall i \in \absy{dom}(T_2)\mathpunct. T_1(i) \sqsubseteq T_2(i)}{\absy{prod}(T_1) \sqsubseteq \absy{prod}(T_2)}
        &&
        \infrule{\forall i \in \absy{dom}(T_1)\mathpunct. T_1(i) \sqsubseteq T_2(i)}{\absy{sum}(T_1) \sqsubseteq \absy{sum}(T_2)}
      \end{align*}
      \vspace{0pt}
    \end{minipage}}
  \caption{Subtyping Rules}
  \label{fig:subtype}
\end{figure}

\begin{theorem}
  The subtyping relation defined by these rules is extensionally sound:
  \begin{align*}
    t_1 \sqsubseteq t_2 \implies \sem[V]{t_1} \subseteq \sem[V]{t_2}
  \end{align*}
\end{theorem}

By contrast, no attempt has been made to make subtyping extensionally
complete (the converse implication).  In particular, there are empty
types that are not considered subtypes of \NEW{(and hence equivalent
  to)} $\absy{none}$.  It is easy to see that the subtyping relation
is purely structural; there are no nominal type, let alone subtype
declarations.

The subtype relation does indeed give rise to a lattice.  The induced
join and meet operations likewise have straightforward syntax-directed
definitions.

\subsubsection{Function Overloading}

Getting function overloading right in conjunction with rich type
systems, especially with subtyping, is notoriously tricky
\cite{Castagna2000,10.1145/75277.75283}.  \textsf{Alea} makes two
simplifying assumptions:
\begin{enumerate}
\item Functions have a single implementation that is a partial
  function defined on a subset of the untyped universe.  Hence a
  function signature of the form $f : t \to u$ means: ``If $f$ is
  applied to an argument value of type $t$, the result value is
  defined and has type $u$''.  Thus, a function can admit multiple
  sound type signatures, but the result computed from an argument
  value must not depend on the type signature used in its checking.
  Ad-hoc polymorphic functions can still be pieced together from
  independent pieces with \emph{disjoint} domains.
\item Function polymorphism is \emph{compact}: A polymorphic function
  may admit infinitely many valid type signatures of the form
  $f : t \to u$, (such as all instantiations of a type scheme in the
  Hindley--Milner system,) but for any actual, valid call with an
  argument of type $t'$, there are only finitely many signatures
  $f : t_i \to u_i$ with $t' \sqsubseteq t_i$ to consider.  The
  effective actual result type is then the greatest lower bound of the
  formal result types, $u' = \bigsqcap u_i$, since they all apply as
  simultaneous guarantees.  Then we write $\vdash f : t' \to u'$ in
  Figure~\ref{fig:typing} below.  The function call is invalid, if the
  set of matching signatures is empty.  How polymorphic type
  signatures are declared is out of scope here.
\end{enumerate}

The former assumption has consequences for \emph{homomorphic}
overloading \cite{Shafarenko2004}: For example, rational (exact)
division and integer (rounding) division must be considered distinct
functions, since they act differently on a pair of integers.  The
former is described by a single type signature
$\mathbb{Q} \times \mathbb{Q} \to \mathbb{Q}$, whereas for the latter
it is useful to give two type signatures,
$\mathbb{Z} \times \mathbb{Z} \to \mathbb{Z}$ and
$\mathbb{N} \times \mathbb{N} \to \mathbb{N}$.\footnote{For clarity,
  we use the traditional notation for function signatures here rather
  than the cumbersome abstract type syntax.}

The latter assumption allows for \emph{parametric} polymorphism in a
general and abstract form, without specifying how it is resolved.  For
example, the addition operator $+$ is highly overloaded:
\begin{itemize}
\item It acts as the usual arithmetic operation with the signatures
  $\mathbb{Q} \times \mathbb{Q} \to \mathbb{Q}$,
  $\mathbb{Z} \times \mathbb{Z} \to \mathbb{Z}$, and
  $\mathbb{N} \times \mathbb{N} \to \mathbb{N}$.  Note that the latter
  implies that the sum of Booleans is a natural number.
\item It acts as the concatenation operation on lists.  This admits an
  infinite parametric family of valid type signatures, namely all of
  the form
  \begin{align*}
    \absy{coll}(\absy{list}, m_1, t_1) \times 
    \absy{coll}(\absy{list}, m_2, t_2) \to 
    \absy{coll}(\absy{list}, m_3, t_3)
  \end{align*}
  with one of $m_1, m_2 \sqsubseteq m_3$ and both
    $t_1, t_2 \sqsubseteq t_3$.  However, for an actual call it
  suffices to consider the single signature with
  $m_3 = m_1 \sqcap m_2$ and $t_3 = t_1 \sqcup t_2$.
\item It acts analogously as the disjoint union on bags.
\item All of the preceding cases form semigroups with some neutral
  element, and all except list concatenation are commutative.  Hence
  they can be used in \emph{reduce} operations on many collection
  types, e.g., a bag of numbers, a list of lists, a list of bags, but
  not a bag of lists (because list concatenation is not commutative).
\end{itemize}

\section{Program Evaluation}
\label{eval}

In this section we shall outline the semantics of \textsf{Alea}
programs.  By nature of the language, there is no fundamental
difference between evaluation of deterministic expressions, static
analysis of outcome distributions, and pseudo-randomized simulation.
All three processes share the same basic syntax-directed big-step
semantics rules, and are constructive enough to be implemented
directly as interpreters.  The difference is in the choice of the
monad that encapsulates results; namely the identity monad for
determinism, the distribution monad for stochastics and a state monad
with a pseudo-random generator for simulation, respectively.

\subsection{Abstract Syntax}

The internal representation of \textsf{Alea} expressions, on which all
evaluation is based, is specified by the following abstract syntax:
\begin{align*}
  \absy{Expr} ::= {}& \absy{var}(\absy{VarId}) \mid
  \absy{const}(\absy{Val} \times \absy{Type}) \mid
  \absy{app}(\absy{FunId} \times \absy{Expr})
  \\
  \mid {}& \absy{choose}(\mathcal{D}(\absy{Expr})) \mid
  \absy{exp}(\absy{Expr}) \mid \absy{dist}(\absy{DistId} \times \absy{Expr})
  \\
  \mid {}& \absy{let}(\absy{Expr} \times \absy{VarId} \times
  \absy{Expr})
  \\
  \mid {}& \absy{nswitch}\bigl(\absy{Expr} \times (\sem[V]{\absy{num}(\absy{rat})} \uplus \{
  \absy{default} \} \nrightarrow \absy{Expr})\bigr)
  \\
  \mid {}& \absy{iter}(\absy{Expr} \times \absy{VarId} \times
  \absy{Expr})
  \\
  \mid {}& \absy{tuple}(\absy{FieldId} \nrightarrow \absy{Expr}) \mid
  \absy{select}(\absy{Expr} \times \absy{FieldId})
  \\
  \mid {}& \absy{cons}(\absy{CaseId} \times \absy{Expr}) \mid
  \absy{cswitch}\bigl(\absy{Expr} \times (\absy{CaseId} \nrightarrow
  \absy{VarId} \times \absy{Expr})\bigr)
\end{align*}

All front-end notation described in Section~\ref{notation} can be
translated to this form; the technical details are out of scope here.
\begin{itemize}
\item The basic term constructs \absy{var}, \absy{const} and
  \absy{app} denote variable references, constants, and
  (deterministic) function applications, respectively.  Note that the
  type of a constant must be specified; in practice it is inferred
  from the form of a literal.  All built-in operators are considered
  functions in this sense.
\item The probability-related constructs \absy{choice}, \absy{exp} and
  \absy{dist} denote fixed distribution, expectation, and drawing from
  a named distribution, respectively.  Note that \absy{dist} takes an
  \absy{Expr}, thus the distribution parameters can be chosen
  dynamically.  \NEW{An expression that does not contain any of these
    constructs is called \emph{syntactically deterministic}.}
\item The scoping-related constructs \absy{let} and \absy{iter} denote
  the binding of a single value and the \emph{flatMap} operation on
  collections of any shape, respectively.
\item The branching-related construct \absy{nswitch} is modeled after
  the \texttt{switch} of the C language family.
\item The product-related constructs \absy{tuple} and \absy{select}
  produce and consume record values, respectively.
\item The sum-related constructs \absy{cons} and \absy{cswitch}
  produce and consume tagged values, respectively.  \absy{cswitch} is
  modeled after pattern matching of functional languages, such as
  \texttt{case}-\texttt{of} in Haskell.
\end{itemize}

\subsection{Type Assignment}

There are no type specifiers in \textsf{Alea} end-user code.  All
relevant type information is inferred in a straightforward bottom-up
procedure.  The rules are depicted in Figure~\ref{fig:typing}.  These
deduction rules produce unique types, hence they inductively define a partial
\emph{type assignment} function:
\begin{align*}
  \sem[T]{e}(\Gamma) =
  \begin{cases}
    t & \text{if}\enspace \Gamma \vdash e : t
    \\
    \text{undefined} & \text{if no such~}t
  \end{cases}
\end{align*}

\begin{figure}
  \centering
  \fbox{\begin{minipage}{.95\textwidth}
  \begin{gather*}
    \infrule{\Gamma(x) = t}{\Gamma \vdash \absy{var}(x) : t}
    \qquad
    \infrule{v \in \sem[V]{t}}{\Gamma \vdash \absy{const}(v, t) : t}
    \qquad
    \infrule{\Gamma \vdash e : t \qquad \vdash f : t \to u}{\Gamma \vdash \absy{app}(f, e) : u}
    \\[\medskipamount]
    \infrule{\bigwedge_{k=1}^n \Gamma \vdash e_k : t_k}{\Gamma \vdash \absy{choose}\bigl( \{e_1 \shortmapsto p_1, \dots, e_n \shortmapsto p_n \}\bigr) : \bigsqcup_{k=1}^n t_k}
    \\[\medskipamount]
    \infrule{\Gamma \vdash e : t \qquad t \sqsubseteq \absy{num}(\absy{rat})}{\Gamma \vdash \absy{exp}(e) : \absy{num}(\absy{rat})}
    \qquad
    \infrule{\Gamma \vdash e : t \qquad \vdash f : t \to u}{\Gamma \vdash \absy{dist}(f, e) : u}
    \\[\medskipamount]
    \infrule{\Gamma \vdash e : t \qquad \Gamma \oplus \{ x \shortmapsto t\} \vdash e' : t'}{\Gamma \vdash \absy{let}(e, x, e') : t'} \qquad
    \\[\medskipamount]
    \infrule{\Gamma \vdash e_0 : t \qquad t \sqsubseteq \absy{num}(\absy{rat}) \qquad \vdash t \bullet C = \{ e_1, \dots, e_k\} \qquad \bigwedge_{k=1}^n\Gamma \vdash e_k : u_k}{\Gamma \vdash \absy{nswitch}(e_0, C) : \bigsqcup_{k=1}^n u_k}
    \\[\medskipamount]
    \infrule{\Gamma \vdash e : \absy{coll}(s, m, t) \qquad s \sqsubseteq s' \qquad \Gamma \oplus \{ x \shortmapsto t \} \vdash e' : \absy{coll}(s', m', t')}{\Gamma \vdash \absy{iter}(e, x, e') : \absy{coll}(s', m \sqcup m', t')}
    \\[\medskipamount]
    \infrule{\bigwedge_{k=1}^n \Gamma \vdash e_k : t_k}{\Gamma \vdash \absy{tuple}\bigl(\{ i_1 \shortmapsto e_1, \dots, i_n \shortmapsto e_n\}\bigr) : \absy{prod}\bigl(\{ i_1 \shortmapsto t_1, \dots, i_n \shortmapsto t_n \}\bigr)}
    \\[\medskipamount]
    \infrule{\Gamma \vdash e : \absy{prod}(T) \qquad T(i) = t}{\Gamma \vdash \absy{select}(e, i) : t}
    \qquad
    \infrule{\Gamma \vdash e : t}{\Gamma \vdash \absy{cons}(i, e) : \absy{sum}\bigl(\{i \shortmapsto t\}\bigr)}
    \\[\medskipamount]
    \infrule{\Gamma \vdash e_0 : \absy{sum}\bigl(\{ i_1 \shortmapsto t_1, \dots, i_n \shortmapsto t_n \}\bigr) \qquad \bigwedge_{k=1}^n \Gamma \oplus \{ x_k \shortmapsto t_k \} \vdash e_k : u_k}{\Gamma \vdash \absy{switch}\bigl(e_0, C \oplus \{ i_1 \shortmapsto (x_1, e_1), \dots, i_n \shortmapsto (x_n, e_n) \}\bigr) : \bigsqcup_{k=1}^n u_k}
  \end{gather*}
  \begin{gather*}
    \infrule{\sem[V]{t} \subseteq \operatorname{dom}(C)}{\vdash t \bullet C = C\bigl(\sem[V]{t}\bigr)}
    \qquad
    \infrule{\sem[V]{t} \not\subseteq \operatorname{dom}(C) \qquad \absy{default} \in \operatorname{dom}(C)}{\vdash t \bullet C = C\bigl(\sem[V]{t} \cup \{ \absy{default} \}\bigr)}
  \end{gather*}
  \vspace{0pt}
    \end{minipage}}
  \caption{Type Inference Rules}
  \label{fig:typing}
\end{figure}

\subsection{Deterministic Evaluation}

The rules for deterministic evaluation, which forms the structural
basis for other interpretations, are depicted in
Figure~\ref{fig:eval-det}.  Note that there are no rules for
evaluating the stochastic constructs $\absy{choose}$, $\absy{exp}$ and
$\absy{dist}$; \NEW{thus the semantics applies to syntactically
  deterministic expressions only.  This will be remedied by the
  following subsections.}  The deduction rules produce unique results,
hence they inductively define a partial \emph{evaluation} function:
\begin{align*}
  \sem[V]{e}(E) =
  \begin{cases}
    v & \text{if}\enspace E \vdash e \leadsto v
    \\
    \text{undefined} & \text{if no such~}v
  \end{cases}
\end{align*}

\begin{figure}
  \centering
  \fbox{\begin{minipage}{.9\textwidth}
      \begin{gather*}
        \infrule{E(x) = v}{E \vdash \absy{var}(x) \leadsto v}
        \qquad
        \infrule{}{E \vdash \absy{const}(v, t) \leadsto v}
        \qquad
        \infrule{E \vdash e \leadsto v \qquad \sem[F]{f}(v) = v'}{E \vdash \absy{app}(f, e) \leadsto v'}
        \\[\medskipamount]
        \infrule{E \vdash e \leadsto v \qquad E \oplus \{ x \shortmapsto v \} \vdash e' \leadsto v'}{E \vdash \absy{let}(e, x, e') \leadsto v'}
        \\[\medskipamount]
        \infrule{E \vdash e_0 \leadsto v_0 \qquad E \vdash C'(v_0) \leadsto v}{E \vdash \absy{nswitch}(e_0, C) \leadsto v}
        \qquad\text{where}\enspace C'(v) =
        \begin{cases}
          C(v) & \text{if defined}
          \\
          C(\absy{default}) & \text{otherwise}
        \end{cases}
        \\[\medskipamount]
        \infrule{E \vdash e \leadsto \absy{the}S(v_1, \dots, v_n) \qquad \bigwedge_{k=1}^n E \oplus \{ x \shortmapsto v_k \} \vdash e' \leadsto v_k'}{E \vdash \absy{iter}(e, x, e') \leadsto v'_1 \oplus \dots \oplus v'_n}
        \\[\medskipamount]
        \infrule{\bigwedge_{k=1}^n E \vdash e_k \leadsto v_k}{E \vdash \absy{tuple}\bigl(\{ i_1 \shortmapsto e_1, \dots, i_n \shortmapsto e_n \}\bigr) \leadsto \{ i_1 \shortmapsto v_1, \dots, i_n \shortmapsto v_n\}}
        \\[\medskipamount]
        \infrule{E \vdash e \leadsto \absy{record}(V) \qquad V(i) = v}{E \vdash \absy{select}(e, i) \leadsto v}
        \qquad
        \infrule{E \vdash e \leadsto v}{E \vdash \absy{cons}(i, e) \leadsto \absy{tag}(i, v)}
        \\[\medskipamount]
        \infrule{E \vdash e_0 \leadsto (i, v_0) \qquad C(i) = (x, e) \qquad E \oplus \{ x \shortmapsto v_0 \} \vdash e \leadsto v}{E \vdash \absy{cswitch}(e_0, C) \leadsto v}
      \end{gather*}
      \vspace{0pt}
    \end{minipage}}
  \caption{Deterministic Big-Step Evaluation Semantics}
  \label{fig:eval-det}
\end{figure}

\begin{ENEW}
  
  \begin{theorem}
    Deterministic evaluation respects types and terminates successfully:
    \begin{align*}
      \left.
        \begin{aligned}
          \sem[T]{e}\bigl(\{ x_1 \shortmapsto t_1, \dots, x_n \shortmapsto t_n \}\bigr) = u
          \\
          \lnot \operatorname{empty} u
          \\
          v_1 \in \sem[V]{t_1}; \dots; v_n \in \sem[V]{t_n}
          \\
          \operatorname{s-deterministic} e
        \end{aligned}
      \right\} \implies \sem[V]{e}\bigl(\{ x_1 \shortmapsto v_1, \dots, x_n \shortmapsto v_n\}\bigr) \in \sem[V]{u}
    \end{align*}
  \end{theorem}

  \begin{proof}
    By induction.
  \end{proof}
\end{ENEW}

\subsection{Stochastic Evaluation}

The rules for stochastic static evaluation are given partially in
Figure~\ref{fig:eval-dist}.  The rules for $\absy{const}$ and
$\absy{let}$ have been spelled out, in order to demonstrate that they
can be derived from the corresponding deterministic rules by
sprinkling in the appropriate natural transformations of the
distribution monad; $\delta$ and $\mu$ are used explicitly, whereas
functorial $\mathcal{D}$ is implied.  In addition, the rules for the
three properly stochastic constructs $\absy{choose}$, $\absy{exp}$ and
$\absy{dist}$ are given.  The deduction rules produce unique results,
hence they define a partial \emph{distribution} function:
\begin{align*}
  \sem[D]{e}(E) =
  \begin{cases}
    P & \text{if}\enspace E \vdash e \rightharpoonup P
    \\
    \text{undefined} & \text{if no such~}P
  \end{cases}
\end{align*}

\begin{figure}
  \centering
  \fbox{\begin{minipage}{.9\textwidth}
      \begin{gather*}
        \infrule{}{E \vdash \absy{const}(v, t) \rightharpoonup \delta(v)}
        \\[\medskipamount]
        \infrule{\bigwedge_{k=1}^n E \vdash e_k \rightharpoonup Q_k}{E \vdash \absy{choose}\bigl(\{ e_1 \shortmapsto p_1, \dots, e_n \shortmapsto p_n \}\bigr) \rightharpoonup \mu\bigl( \{ Q_1 \shortmapsto p_1, \dots, Q_n \shortmapsto p_n \} \bigr)}
        \\[\medskipamount]
        \infrule{E \vdash e \rightharpoonup \{ \absy{const}(x_1) \shortmapsto p_1, \dots, \absy{const}(x_n) \shortmapsto p_n\} \qquad m = \sum_{k=1}^n p_n \cdot x_n}{E \vdash \absy{exp}(e) \rightharpoonup \delta(\absy{const}(m))}
        \\[\medskipamount]
        \infrule{E \vdash e \rightharpoonup P = \{ v_1 \shortmapsto p_1, \dots, v_n \shortmapsto p_n \} \qquad \bigwedge_{k=1}^n \sem[F]{f}(v_k) = Q_k}{E \vdash \absy{dist}(f, e) \rightharpoonup \mu\bigl( \{ Q_1 \shortmapsto p_1, \dots, Q_n \shortmapsto p_n \} \bigr)}
        \\[\medskipamount]
        \infrule{E \vdash e \rightharpoonup P = \{ v_1 \shortmapsto p_1, \dots, v_n \shortmapsto p_n \} \qquad \bigwedge_{k=1}^n E \oplus \{ x \shortmapsto v_k \} \vdash e' \rightharpoonup Q_k}{E \vdash \absy{let}(e, x, e') \rightharpoonup \mu\bigl( \{ Q_1 \shortmapsto p_1, \dots, Q_n \shortmapsto p_n \} \bigr)}
      \end{gather*}
      \vspace{0pt}
    \end{minipage}}
  \caption{Stochastic Big-Step Evaluation Semantics (Excerpt)}
  \label{fig:eval-dist}
\end{figure}

\begin{ENEW}
  We have only given enough stochastic evaluation rules such that the
  reader can extrapolate to a full system, given some working
  knowledge how monads are employed in denotational semantics.  Thus,
  given the incomplete material presented here, no fully formal
  properties can be stated.  In order to convey the logical intention
  nevertheless, the following and subsequent similar theorems are
  marked with (Sketch), to express that the underlying definitions are
  to be completed in a way that makes the theorem hold eventually.

  \begin{theorem}[Sketch]
    Stochastic evaluation respects types and terminates successfully:
    \begin{align*}
      \left.
        \begin{aligned}
          \sem[T]{e}\bigl(\{ x_1 \shortmapsto t_1, \dots, x_n \shortmapsto t_n \}\bigr) = u
          \\
          \lnot \operatorname{empty} u
          \\
          v_1 \in \sem[V]{t_1}; \dots; v_n \in \sem[V]{t_n}
        \end{aligned}
      \right\} \implies \sem[D]{e}\bigl(\{ x_1 \shortmapsto v_1, \dots, x_n \shortmapsto v_n\}\bigr) \in \mathcal{D}\bigl(\sem[V]{u}\bigr)
    \end{align*}
  \end{theorem}

  \pagebreak
  \begin{theorem}[Sketch]
    Stochastic evaluation coincides with deterministic evaluation on
    syntactically deterministic expressions:
    \begin{align*}
      \begin{aligned}
        \operatorname{s-deterministic} e
      \end{aligned}
      \implies \sem[D]{e}(E) = \delta_{\sem[V]{e}(E)}
      \quad\text{(if both defined)}
    \end{align*}
  \end{theorem}
  
\end{ENEW}

\subsection{Pseudo-Randomized Evaluation}

The rules for pseudo-random evaluation are given partially in
Figure~\ref{fig:eval-rand}.  We write
$e \xrightarrow{s \shortrightarrow s'} v$ to say that the evaluation
of $e$, with the random generator being in state $s$, results in value
$v$, with the random generator transitioning to state $s'$.  We assume
that the generator supports an elementary operation,
$\operatorname{random}$, that selects a number in $1, \dots, n$ according
to a specified $n$-tuple of probabilities.

Note that, unlike the preceding rule systems, this semantics is no
longer uniquely valued; for example, the $\absy{tuple}$ rule
does not specify the sequential order in which fields are processed.
\NEW{Thus, the rules define a set-valued evaluation function,
  parameterized also over the initial state:}
\begin{align*}
  \sem[R]{e}(E, s) = \{ v \mid \exists s'\mathpunct. E \vdash e \xrightarrow{s \shortrightarrow s'} v \}
\end{align*}

\begin{figure}
  \centering
  \fbox{\begin{minipage}{.9\textwidth}
      \begin{gather*}
        \infrule{}{E \vdash \absy{const}(v, t) \xrightarrow{s \shortrightarrow s} v}
        \\[\medskipamount]
        \infrule{\bigwedge_{k=1}^n E \vdash e_k \xrightarrow{s_{k-1} \shortrightarrow s_k} v_k}{E \vdash \absy{tuple}\bigl(\{ i_1 \shortmapsto e_1, \dots, i_n \shortmapsto e_n \}\bigr) \xrightarrow{s_0 \shortrightarrow s_n} \{ i_1 \shortmapsto v_1, \dots, i_n \shortmapsto v_n\}}
        \\[\medskipamount]
        \infrule{\operatorname{random}(s; p_1, \dots, p_n) = (s', k) \qquad E \vdash e_k \xrightarrow{s' \shortrightarrow s''} v}{E \vdash \absy{choose}\bigl(\{ e_1 \shortmapsto p_1, \dots, e_n \shortmapsto p_n \}\bigr) \xrightarrow{s \shortrightarrow s''} v}
      \end{gather*}
      \vspace{0pt}
    \end{minipage}}
  \caption{Pseudo-Random Big-Step Evaluation Semantics (Excerpt)}
  \label{fig:eval-rand}
\end{figure}

\begin{ENEW}
  
  \begin{theorem}[Sketch]
    Pseudo-random evaluation respects types and terminates successfully:
    \begin{align*}
      \left.
        \begin{aligned}
          \sem[T]{e}\bigl(\{ x_1 \shortmapsto t_1, \dots, x_n \shortmapsto t_n \}\bigr) = u
          \\
          \lnot \operatorname{empty} u
          \\
          v_1 \in \sem[V]{t_1}; \dots; v_n \in \sem[V]{t_n}
        \end{aligned}
      \right\} \implies
      \emptyset \neq \sem[R]{e}\bigl(\{ x_1 \shortmapsto v_1, \dots, x_n \shortmapsto v_n\}, s\bigr) \subseteq \sem[V]{u}
    \end{align*}
  \end{theorem}

  \begin{theorem}[Sketch]
    Pseudo-random evaluation coincides with deterministic evaluation on
    syntactically deterministic expressions:
    \begin{align*}
      \begin{aligned}
        \operatorname{s-deterministic} e
      \end{aligned}
      \implies \sem[R]{e}(E) = \operatorname{const}_{\sem[V]{e}(E)}
      \quad\text{(if both defined)}
    \end{align*}
  \end{theorem}
  
  The intended relationship between stochastic and pseudo-random
  evaluation should be such that, given an ideal generator, the latter
  is an unbiased sampling procedure for the former.

  \begin{theorem}[Sketch]
    Assume that pseudo-random evaluation is equipped with a generator
    that has the following ideal properties:
    \begin{enumerate}
    \item Values are produced with the prescribed probabilities over
      truly random, equiprobable states $s$:
      \begin{align*}
        \mathcal{P}\bigl(\operatorname{random}(s; p_1, \dots, p_n) = k\bigr) = p_k
      \end{align*}
    \item Outcomes for distinct states are stochastically independent.
    \end{enumerate}

    Then pseudo-random evaluation yields particular results with the
    probabilities calculated by stochastic evaluation over truly
    random, equiprobable states $s$:
    \begin{align*}
      \left.
        \begin{aligned}
          \sem[T]{e}\bigl(\{ x_1 \shortmapsto t_1, \dots, x_n \shortmapsto t_n \}\bigr) = u
          \\
          \lnot \operatorname{empty} u
          \\
          v_1 \in \sem[V]{t_1}; \dots; v_n \in \sem[V]{t_n}
        \end{aligned}
      \right\} \implies
      \mathcal{P}\bigl(\sem[R]{e}(E, s) = v\bigr) = \sem[D]{e}(E)(v)
    \end{align*}
  \end{theorem}

\end{ENEW}

\section{Application Examples}

\NEW{We give \textsf{Alea} implementations of the informal examples
  from section~\ref{intro} and some others, and briefly discuss style
  and practical results.}

\subsection{Introduction Revisited}

A slightly biased instance of the singular Roman coin toss experiment
(Example~\ref{example:denarius}) could be specified like this:
\begin{lstlisting}[gobble=4,xleftmargin=4em]
  % ~bernoulli(0.503) ? @head : @ship
\end{lstlisting}

A specification of the more complex dice-rolling
Example~\ref{example:vampire} is depicted in Figure~\ref{fig:vampire}.
For a discussion of the concepts of notation and how they are used,
see the next subsection.  The probability of winning, by any amount, is
calculated by the \textsf{Alea} interpreter as exactly
$0.90661502737169$.\;\footnote{The interpreter is a reference
  implementation that employs abstract syntax data structures and an
  evaluation procedure that follow the description in
  section\ref{eval} very closely, without sophisticated optimizations.
  The calculation has been measured to take 11.5\,s on a Core
  i7-12700H processor running Ubuntu 24.04.2 and OpenJDK 21.0.6.}

\begin{ENEW}

  \subsection{Case Study}
  
  There is anecdotal evidence that the syntax and semantics of
  \textsf{Alea} are reasonably easy to grasp for non-programmers.  A
  test person with good mental calculation skills, but no training in
  programming or computer science, noticed that the informal statement
  of the rule as paraphrased in Example~\ref{example:vampire} is
  ambiguous about one detail; namely, whether ones that are rolled on
  the extra dice should be counted towards failure.  They
  \begin{itemize}
  \item immediately noted that the affirmative answer to this question
    is located in the expression in line~6,
  \item immediately agreed that the alternative would be expressed
    appropriately by changing the line to
    \begin{lstlisting}[gobble=8,numbers=left,numberstyle=\footnotesize,firstnumber=6,xleftmargin=4em]
      % fails := (+)< d = 1 | d <- dice1 >;
    \end{lstlisting}
  \item estimated that the alternative would improve the odds of the
    player, but only very slightly; the interpreter calculates the
    probability as exactly $0.91182365$.
  \end{itemize}

  Broader studies that are able to evaluate the suitability of
  \textsf{Alea} for various groups of target audience (stochastics
  students, programming students, general laypeople) more objectively
  shall be performed in future work.
  
\end{ENEW}

\begin{figure}
  \begin{lstlisting}[gobble=6,xleftmargin=2em,numbers=left,numberstyle=\footnotesize,frame=single,escapechar=!]
    % dice1   := < ~uniform{1..10} | _ <- <1..7> >;  -- Roll seven ten!-!sided dice.
    % tens    := < d | d <- dice1; d = 10 >;        -- For every die that shows a ten,
    % dice2   := < ~uniform{1..10} | _ <- tens >;    -- roll another
    % dice    := dice1 + dice2;                      -- and add it to the pool.
    % succs   := (+)< d >> 5 | d <- dice >;         -- Count the number of dice that show values
    % fails   := (+)< d = 1 | d <- dice >;         -- greater than five. Compare to the number 
    % diff    := succs - fails;                       -- of dice that show a one.
    % verdict := diff >> 0 ? @win(diff)       -- If the difference is positive, you win by that amount.
    %            : (succs = 0 && fails >> 0    -- If there are no values greater than five but
    %               ? @botch : @lose)          -- there is a one, you lose badly.
  \end{lstlisting}
  \caption{Implementation of Example~\ref{example:vampire} from Section~\ref{intro}}
  \label{fig:vampire}
\end{figure}

\subsection{Bonus Use Case: Yahtzee}

As a real-world application example, we specify the scoring system of
the dice game Yahtzee \cite{yahtzee}.  Here we consider only the
scoring of a single throw of five dice, ignoring the further game
mechanics, namely selective rerolling, the thirteen rounds, and the
bonuses.

The \textsf{Alea} code for the basic scheme is depicted in
Figure~\ref{fig:yahtzee}.  Some comments on the notation and proposed
style:

\begin{figure}
  \begin{lstlisting}[gobble=6,xleftmargin=2em,numbers=left,numberstyle=\footnotesize,frame=single]
    % dice := < ~uniform{1 .. 6} | _ <- <1 .. 5> >;
    % (
    %     Dice:          dice,
    %     Aces:          (+)< d | d <- dice; d = 1 >,
    %     Twos:          (+)< d | d <- dice; d = 2 >,
    %     Threes:        (+)< d | d <- dice; d = 3 >,
    %     Fours:         (+)< d | d <- dice; d = 4 >,
    %     Fives:         (+)< d | d <- dice; d = 5 >,
    %     Sixes:         (+)< d | d <- dice; d = 6 >,
    %     Chance:        (+)(dice),
    %     ThreeofaKind:  (+)(dice) * (max(mults(dice)) >= 3),
    %     FourofaKind:   (+)(dice) * (max(mults(dice)) >= 4),
    %     FullHouse:     25 * (mults(dice) = <2, 3>),
    %     SmallStraight: 30 * (dice >= <1 .. 4> || dice >= <2 .. 5> || dice >= <3 .. 6>),
    %     LargeStraight: 40 * (dice >= <1 .. 5> || dice >= <2 .. 6>),
    %     Yahtzee:       50 * (mults(dice) = <5>)
    % )
  \end{lstlisting}
\caption{Yahtzee Scoring}
  \label{fig:yahtzee}
\end{figure}

\begin{itemize}
\item Line~1 makes five invocations of the uniform distribution on the
  set of numbers $\{1, \dots, 6\}$, and collects the results in a bag.
  The concept of collections of independent, identically distributed
  (iid) random variables is quite ubiquitous; hence it could deserve a
  specific notation, but we demonstrate here that its semantics are
  already subsumed by bag comprehension.
\item Lines 2--17 compute various scores depending on the rolled dice
  values, and store them in a record.
\item Lines 4--9 sum only the dice values that are equal to one, two,
  etc., respectively.
\item Lines 10--12 sum all dice values conditionally.  Lines 13--16
  take fixed values conditionally.  A condition is imposed by
  multiplication with a Boolean expression.
\item Lines 11--13 and 17 make use of the \emph{mults} function, a
  useful statistic that maps a bag of arbitrary values to the bag of
  their multiplicities, e.g., $\langle a, a, a, b, c \rangle$ to
  $\langle 3, 1, 1 \rangle$.
\item Lines 14--15 use the \emph{superbag} relation, written as
  $\geq$.
\end{itemize}

Analysis of the outcomes is possible by subjecting the depicted
expression to further computations.  For example, to find the
probability of obtaining at least 17 points in the category `three of
a kind', take the expectation value of the corresponding Boolean
expression:
\begin{lstlisting}
  E(dice := <...>; (...).ThreeofaKind >= 17)
\end{lstlisting}
The result, computed by the \textsf{Alea} stochastic interpreter, is $17/144$.

\section{Current Implementation State}

A processing tool for \textsf{Alea} programs has been implemented in
\textsf{Java}.  It features a type checker and two interpreters, one
for stochastic analysis and one for pseudo-random evaluation.

While this tool is good enough for first evaluations, for both a
full-fledged stochastics teaching aid and a game support engine many
features are still missing.  For the former, \emph{visualizations} of
the structure expressions and the distributions of outcomes, and
summary diagrams for parameterized experiments would greatly
enhance the experience.  For the latter, a \emph{server}
infrastructure, where game rules can be deployed and queries of
various kind can be answered, is needed.

Apart from such user interfaces, the prospect of \emph{compiling}
\textsf{Alea} code to some other language, while performing
optimizations, has not been explored.

In general, all software related to \textsf{Alea} shall eventually be
released as open source.

\section{Conclusion}

We have presented \textsf{Alea}, a simple functional programming
language geared towards the declarative specification of simple random
experiments.  Due to the particular nature of the application domain,
\textsf{Alea} is both lacking many common features of more general FP
languages, and equipped with specific novel features to suit the
application.  In particular, the bag comprehension notation and the
numeric Booleans stand out as ubiquitously useful.  As a result of the
choice of features and the design of the front-end syntax, the
language has a distinctive elementary mathematics-like feeling, and
should be easy and fun to use by non-experts in programming.

Due to the deliberate design choice to make \textsf{Alea}
Turing-incomplete, plausible use cases where expressive power is
objectively lacking are bound to arise.  Thus, the library of opaquely
predefined functions and distributions is expected to grow, and more
design work concerning language extensions for syntactic sugar and
reusable user-defined subprograms is needed.

Already in fairly simple example code, combinatorial explosion can
arise, in particular in conjunction with \emph{distinguishable} random
elements that are collected in lists.  While this is no practical
problem for pseudo-random simulation at all, the naïve way to compute
distributions, even simple marginal ones, directly from the semantic
rules can get disappointingly slow.  Since \textsf{Alea} is not
referentially transparent with regard to randomness, optimization
techniques such as lazy evaluation or program slicing cannot be
applied straightforwardly.  It appears that stochastic independence
can be exploited to address this issue, but fundamental research into
the matter has only just started.

\section*{Acknowledgments}

Thanks to J.\ Gibbons for helpful discussions on the syntax and
semantics of collections.  Thanks to B.~Hanke for helpful discussion
on the pragmatics of dice rolls.

\raggedright
\bibliographystyle{eptcs}
\bibliography{alea}

\end{document}